\documentclass[aps,preprint]{revtex4}
\usepackage{graphicx}

\begin{document}

\title{Tunable pinning in superconducting films with magnetic micro-loops}

\author{A. V. Silhanek}
\affiliation{INPAC-Institute for Nanoscale Physics and Chemistry, Nanoscale Superconductivity and Magnetism $\&$ Pulsed
Fields Group, K. U. Leuven Celestijnenlaan 200 D, B-3001 Leuven, Belgium.}

\author{V. Metlushko}
\affiliation{Department of Electrical and Computer Engineering, University of Illinois, Chicago, IL 60607.}

\author{W. Gillijns}
\affiliation{INPAC-Institute for Nanoscale Physics and Chemistry, Nanoscale Superconductivity and Magnetism $\&$ Pulsed
Fields Group, K. U. Leuven Celestijnenlaan 200 D, B-3001 Leuven, Belgium.}

\author{B. Ilic}
\affiliation{Cornell Nanofabrication Facility, School of
Applied and Engineering Physics, Cornell University, Ithaca, New
York 14853}

\author{V. V. Moshchalkov}
\affiliation{INPAC-Institute for Nanoscale Physics and Chemistry, Nanoscale Superconductivity and Magnetism $\&$ Pulsed
Fields Group, K. U. Leuven Celestijnenlaan 200 D, B-3001 Leuven, Belgium.}

\date{\today}

\begin{abstract}
We study the flux pinning properties of superconductor/magnetic micro-rings lattice hybrid structures. The used open triangular  micromagnets represent an eight-fold degree of freedom system, with six polarized and two flux-closure possible states. By conveniently choosing the magnetic state of the underlying rings it is possible to induce different pinning potentials. We show that the magnetic vortex state with minimum stray field produces a weaker pinning in comparison with the polarized states.
\end{abstract}

\pacs{PACS numbers: ..............................}

\maketitle

Regular arrays of nanoengineered pinning centers in conventional superconductors have proven to be an efficient way to enhance the critical current $I_c$. In general, the extent to which these centers influence the superconducting properties of the system depends on the details of the used nanostructuring. Nowadays it is possible to delineate a hierarchical list of different pinning sites going from the relatively weak blind holes\cite{raedts}, in-plane and out of plane magnetic dots\cite{lyuksyutov}, to the strong pinning produced by antidots\cite{baert}. For all these structures there is a natural limitation as they generate a nearly immutable pinning potential unable to be modified once defined.

Interestingly, as opposed to singly-connected magnetic dots and non-magnetic pinning structures, magnetic micro-loops can be readily set in several different magnetic states\cite{ab}. Every of these magnetic states interacts differently with the vortices in the superconductor thus allowing one to control the easy direction of the vortex drift. It is particularly this remarkable flexibility to manipulate the vortex motion which makes this kind of pinning potentials attractive for practical applications although still a clear experimental corroboration is pending.

In this work we explore tunable pinning centers generated by an array of triangular rings and demonstrate that by changing the magnetic state of the underlying micro-loops it is possible to modify the pinning properties of the superconductor. In particular we show that the flux-closure magnetic vortex state represents a less efficient pinning potential compared with the polarized state.

The used sample consists of a closely packed square array of cobalt equilateral triangular loops (250 nm wide, 23 nm thick, with lateral size $d=$ 2 $\mu$m and separated by $s=$ 250 nm from the neighboring elements) fabricated by electron-beam lithography and lift-off on a silicon wafer. A scanning electron microscope (SEM) image of the array is shown in Fig.~\ref{fig1}(a). We evaporated a 5 nm thick Ge layer on top to avoid proximity effects\cite{buzdin-RMP} and then a superconducting Pb film (thickness $t$ = 25 nm, critical temperature $T_c$ = 7.220 K, residual resistivity ratio RRR $\sim$ 15 and electronic mean free path\cite{meanfreepath} $\ell \sim$ 40 nm) which is covered with a protective 20 nm thick Ge capping layer. In order to ensure an homogeneous current distribution, the continuous Ge/Pb/Ge trilayer is patterned into a transport bridge by optical lithography and ion-mill etching. The 400 $\mu$m wide transport bridge is aligned along $-{\bf x}$ direction as schematically indicated in Fig.~\ref{fig1}(a). In this way the Lorentz force lies along the higher symmetry axis of the underlaying array. The vortex pinning properties of this hybrid sample were investigated by electrical transport measurements using a Quantum Design cryostat with conventional electronics.

It is worth noticing that if solid singly-connected triangles are used as in Ref.[\onlinecite{villegas}] a flux-closure state with a strong z-component in the middle of the triangle is formed\cite{ab}. In contrast to this, ring structures favor a flux-closure state because the highly energetic vortex core is removed\cite{comment1}. Inserts (b) and (c) in Fig.~\ref{fig1} show two typical magnetic states of triangular rings elements measured by Magnetic Force Microscopy (MFM). If a magnetic field $H$ higher than the saturation value is applied along one of the sides of the triangle, the element is magnetized into a single domain state with in-plane magnetization. A full magnetization loop $M(H)$ at $T=10$ K for applied field ${\bf H}$ along ${\bf y}$ direction as determined by SQUID measurements is shown in Fig.~\ref{fig1}(d)\cite{Tcomment}.

When a large magnetic field higher than saturation field is applied parallel to the edge direction, the ring is magnetized into a single domain state. The domain walls are formed as the magnetic field is reduced to zero. The exact magnetic patterns depend on the magnetic history.
When $H$ is reversed to about -43 mT and then reduced to zero (open symbols in Fig.~\ref{fig1}(d)), domain walls are formed as shown in Fig.~\ref{fig1}(b). At the three different corners the magnetic moment gradually rotates by 120$^{\circ}$\cite{vitali-IEEE}. However, if starting from the single domain fully polarized state $H$ is directly reduced to zero a different domain pattern is observed (see Fig.~\ref{fig1}(c)). Now at the lower right corner of the ring, the magnetic moment gradually rotates by 120$^{\circ}$ whereas at the other two corners the magnetic moments form either transverse or vortex type domain walls. In this case the stray field contrast of the z-component, indicated in the inserts as white and black spots, is much stronger than for the 120$^{\circ}$ walls. Object Oriented Micromagnetic Simulations (OOMMF) suggest that each of the three segments of the triangular shaped rings are nearly single domain states, as indicated by the arrows in Fig.~\ref{fig1}(b) and (c). Different orientation of these single domain states form different domain patterns. As a result the triangular shaped rings have two vortex states of opposite chirality, and six polarized states. The type and orientation of the elements can be controlled by a small external in-plane magnetic field. For detailed description of magnetization process in triangular ring we refer readers to Ref.\cite{vitali-IEEE} and \cite{vavassori}.

It has been previously shown that for pinning purposes only the out-of-plane component of the stray field counts\cite{lieve}. This effect suggests
that in our case the pinning efficiency can be tuned from weak to strong limits by simply switching from flux-closure to the
polarized magnetic state of the Co loops. In order to check whether the magnetic state of the underlying micromagnets indeed influences the superconducting properties of the thin film we first determined the normal-superconductor (N/SC) phase boundary. Fig.~\ref{fig:2}(a) shows the N/SC transition as estimated by 10$\%$ of the normal state resistance $R_n$ for two of the magnetic states and compare it with the results for a simultaneously grown plain film. The plain film exhibits a featureless nearly linear upper critical field $H_{c2}(T)$ from which is possible to estimate a superconducting coherence length $\xi(0)=$33$\pm$3 nm. In the case of the Pb film with the array of triangles underneath, a parabolic
background is observed around $H=0$ regardless the initial magnetic state. A similar behavior has been reported for F/S bilayers\cite{lange} and samples with square arrays of holes\cite{rosseel}. In our sample this transition from linear to parabolic field dependence occurs at $T_o \approx$ 7.16 K where 1.8$\xi(T_o) \sim$ 1.26 $\mu$m exceeds the size $w$ of the domains where superconductivity first nucleates\cite{rosseel}. For higher temperatures the phase boundary can be approximated by\cite{VVM} $T_c(H)/T_c(0)=1-(\alpha H)^2$, with $\alpha=\xi(0)\pi w /2\sqrt 3 \phi_0$ and $\phi_0$ the flux quantum. Using this expression to fit the data in Fig.~\ref{fig:2}(a) we obtain $w \sim$ 1.26 $\mu$m which agrees  with the estimation obtained above from the coherence length.

From Fig.~\ref{fig:2}(a) it is apparent that the patterned sample exhibits clear commensurability features at $nH_1$, with $n=\frac{1}{2}, 1, \frac{3}{2},$ 2 and $H_1=\phi_0/((dcos(\pi/6)+s)(d+s))$ = 0.465 mT (the unit cell is shown schematically in Fig.~\ref{fig1}(a)). Furthermore, a
direct comparison of the phase boundary for the magnetic vortex state with the magnetized state indicates that matching features, and therefore collective effects, are far less prominent in the flux-closure vortex state. This effect is a direct consequence of the different stray field generated by these two magnetic states.

Although the flux line lattice configuration at the main matching fields is expected to follow the rectangular symmetry of the underlying magnetic landscape\cite{stoll}, it is not trivial to determine where actually the vortices locate within the unit cell. In the magnetized state two different scenarios may arise as to whether the stray field associated with the in-plane dipole can generate a vortex-antivortex pair or not. If there is no
vortex-antivortex pair, then flux lines sit on top of the positive z-component of the stray field otherwise flux lines will be attracted by the antivortices generated by the stray field and thus sit close to the maximum negative z-component.

In the magnetic vortex state, the three identical domain walls at the vertices of the triangles do not impose any preferential pinning for an isolated element. However, if $\xi(T) > s$ this three fold degeneracy is lifted by the close proximity between two adjacent triangles and
vortices sit where the stray field maximizes. This is consistent with the fact that matching features in all cases are visible only for $T>$ 7.16 K where $\xi(T) >$ 0.45 $\mu$m $> s$.

Lets now focus on the flux pinning properties of the system. To that end we measured the critical current $I_c(H)$ at $T/T_c=0.994$ using a voltage criterion of $100 \mu$V for three contrasting magnetic states as shown in Fig.~\ref{fig:2}(b). The ${\bf
m}=\frac{1}{2}(\sqrt 3{\bf x}+{\bf y})$ magnetized state is built by applying an in-plane field 150$^{\circ}$ off the current direction whereas for the ${\bf m}={\bf y}$ state the in-plane field is applied perpendicular to the current direction. In the ${\bf m}={\bf y}$ state the pinning force is maximum and therefore an optimal pinning efficiency is expected in agreement with the observed behavior. At the commensurability fields, the vortex lattice is very stable (elastic constants have a local maximum) and the difference between the two magnetic states becomes negligible. The most striking behavior however appears when the magnetic triangles are set in the vortex state. Here, {\it the critical current is substantially reduced and only commensurability at $H_1$ remains}. A similar result has been reported by Van Look {\it et al}.\cite{lieve}
for an array of magnetic dots although in that work a much more complicated demagnetization protocol was followed. The lack of features at rational matching fields in the flux-closure state can be ascribed to the weaker pinning potential which favors a more disordered flux line lattice. At lower temperatures (not shown) a kink in $I_c(H)$ at $H=2H_1$ is still visible although less pronounced than for the polarized magnetic rings.

To summarize, we have explored the dynamic pinning properties of a conventional superconductor on top of an array of micromagnets composed by Co open triangles. We show that the system studied represents a very flexible and controllable flux pinning structure. The used triangular shape of the magnets gave us full control of the chirality of the flux-closure state and the possibility to induce six different dipolar states by following very simple magnetization protocols. A clear contrast in the vortex pinning properties of the flux-closure and the polarized states is observed.

V.M. acknowledge funding support from U.S. NSF, grants ECS-0202780 and DMR-0210519. A.V.S. is grateful for the support from the FWO-Vlaanderen. This work was also partially supported by the GOA project.

\newpage


\newpage

\begin{figure}[htb]
\centering
\caption{(Color online) (a) SEM image indicating the used spatial coordinates and the unit cell of the pinning array, (b) magnetic vortex and (c) polarized states as seen by MFM. (d) Magnetic hysteresis loop $M(H)$ of the Co-rings array at $T=10$ K for ${\bf H} \parallel {\bf y}$ normalized to its saturation value $M_s=911$ emu/cm$^3$. Open symbols correspond to the flux-closure state.} \label{fig1}
\end{figure}

\begin{figure}[h!]
\centering
\caption{(Color online) (a) N/SC phase boundary estimated by 10$\% R_n$ criterion for a reference plain film, and a film on top of Co rings array in the flux-closure and polarized states. (b) critical current estimated by 100$\mu$V criterion at $T/T_c=$0.994 for the magnetic closure state and the ${\bf m}=+{\bf y}$ and ${\bf m}=\frac{1}{2}(\sqrt 3{\bf x}+{\bf y})$ polarized states.}
\label{fig:2}
\end{figure}


\begin{thebibliography}{10}
\bibitem{comment1}
We have found experimentally that for separations between adjacent triangles larger than 100 nm no magnetic interaction is expected and thus each triangle can be regarded as an isolated element.
\bibitem{Tcomment}
Almost no temperature dependence of the magnetization was observed for $10$ K $< T < 300$ K.
\end{thebibliography}
\end{document}